\documentclass[12pt]{iopart}
\usepackage{lineno}

\usepackage{epsfig}
\usepackage{wrapfig}
\usepackage{graphicx}
\usepackage{amsfonts}
\usepackage{amssymb}
\usepackage{subfigure}
\usepackage{helvet}
\usepackage{hhline}
\usepackage{appendix}
\usepackage{multirow}
\usepackage[plainpages=false, linktocpage=true]{hyperref}
\usepackage{color}
\usepackage{tabularx}

\newcommand \phe{$\phi\rightarrow e^{+}e^{-}$ }
\newcommand \omge{$\omge\rightarrow e^{+}e^{-}$ }
\newcommand \phk{$\phi\rightarrow K^{+}K^{-}$ }

\newcommand \sqn{$\sqrt{s_{_{NN}}}$ }

\newcommand \dau{$~d+Au$ }
\newcommand \au{$Au+Au$ }
\newcommand \pp{$p+p$ }
\newcommand{\degree}{\ensuremath{^\circ}}
\newcommand{\cerenkov}{$\check{C}$erenkov }
\newcommand{\npid}{``No Kaon PID''}
\newcommand{\opid}{``One Kaon PID''}
\newcommand{\tpid}{``Two Kaons PID''}

\begin{document}

\title[Proceedings for SQM 2008]{$\phi$- meson Production at RHIC energies using the PHENIX Detector}

\author{Deepali Sharma (For the PHENIX Collaboration)}

\address{Department of Particle Physics, The Weizmann Institute of
  Science, Rehovot, 76100, Israel}

\ead{deepali.sharma@weizmnann.ac.il}

\begin{abstract}
Light vector mesons are among the most informative probes to
understand the strongly coupled Quark Gluon Plasma created at RHIC.
The suppression of light mesons at high transverse momentum,
compared to expectations from scaled $p+p$ results, reflects the
properties of the strongly interacting matter formed. The
$\phi$-meson is one of the probes whose systematic measurement in
$p+p$, $d+Au$ and $Au+Au$ collisions can provide useful
information about initial and final state effects on particle
production. The mass, width and branching ratio of the $\phi$-meson
decay in the di-kaon and di-electron decay channels could be modified
in \au collisions due to the restoration of chiral symmetry in the QGP. 

The PHENIX experiment at RHIC has measured $\phi$-meson production
in various systems ranging form $p+p$, $d+Au$ to $Au+Au$ collisions
via both its di-electron and di-kaon decay modes. A summary of PHENIX
results on invariant spectra, nuclear modification factor and elliptic
flow of the $\phi$-meson are presented here. 

 
\end{abstract}

\maketitle

\section{Introduction}
The $\phi$- meson plays a unique role in the study of the hot
and dense medium created in relativistic heavy-ion collisions. It is
is the lightest bound state of hidden strangeness $s\overline{s}$, has a
small interaction with other non-strange hadrons and hence carries
information from the early partonic stages of the system
evolution. Comparing the elliptic flow ($v_{2}$) of $\phi$ to the
$v_{2}$ of other multistrange hadrons ($\Omega$ and $\Xi$) or
particles composed of lighter quarks ($u$ and $d$) or heavier charm
quark, provides information about the partonic collectivity of the
medium.

Furthermore the $\phi$ can provide important information on particle
production mechanisms, since it is a meson but has a mass similar
to $p$ and $\Lambda$ baryons. The measurement of its nuclear
modification factor, $R_{AA}$ adds to the picture of particle 
suppression and its dependence on particle mass and composition
supporting hydrodynamics and recombination models. 

The $\phi$ can also be sensitive to the restoration of chiral
symmetry. A certain fraction of the $\phi$ can decay inside the hot
and dense media leading to a change in its spectral
function\cite{ref1, ref2}. This modification can be seen by studying
the low-momentum $\phi$ decaying inside the media and reconstructed
via the di-electron decay channel. Since leptons are not subject to
the strong interaction, they preserve their production information. A
change in mass or width ($\Gamma$) of $\phi$ inside the medium can
lead to a change in the relative branching ratios of the $\phi \rightarrow
K^{+}K^{-}$ and $\phi \rightarrow e^{+}e^{-}$ decay modes. Since
$m_{\phi} \approx 2\times m_{K}$, small changes in $\phi$ or $K$ can
induce significant changes in the branching ratio.
\vspace{-0.3cm}
\section{The PHENIX Detector }

\begin{wrapfigure}{r}{0.5\textwidth}
  \vspace{-1.5cm}
  \begin{center}
    \includegraphics[width=0.4\textheight]{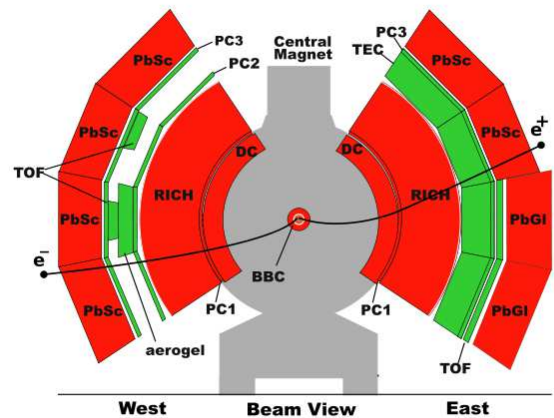}
  \end{center}
  \vspace{-1.5cm}
  \caption{Schematic view of the PHENIX experiment, highlighting the
    subsystems used in di-electron analysis.}
  \label{fig:fig_phenix}
\end{wrapfigure}


The PHENIX detector\cite{phnx} at RHIC (Relativistic Heavy Ion
Collider) has been designed to measure both leptons and hadrons. A
schematic view of the PHENIX detector is shown in
Fig.~\ref{fig:fig_phenix}. Each of the two central arm spectrometers
covers 90\degree in azimuth and $\pm$0.35 in pseudorapidity and has
the capability to measure neutral and charged particles. The
high-resolution multi-wire proportional Drift Chambers (DC) together
with the first layer of Pad Chambers (PC1) provide the charged
particle tracking and momentum measurement. The typical momentum
resolution is $\sigma (p_{T})/p_{T} \approx 1.0 \% p_{T} 
\oplus $ 1.1$\%$. The Kaons are identified by using the timing
information from a high resolution Time of Flight (TOF) detector and
the Lead Scintillator (PbSc) part of the Electromagnetic Calorimeter
(EMCal), with good $\pi/K$ separation over the momentum range 0.3 -
2.5 GeV/\textit{c} and 0.3 - 1 GeV/\textit{c}, respectively. The electrons are
identified using a Ring Imaging \cerenkov Detector (RICH) and by
requiring the energy measured in the EMCal to match the measured momentum
of the charged tracks in the DC. The Zero Degree Claorimeters (ZDC's)
and Beam Beam Counters (BBC's) are dedicated subsystems that measure
global quantities such as luminosity, collision vertex and event
centrality. The minimum bias trigger is derived by a coincidence
between the two BBCs; in $p+p$ and $d+Au$ the trigger requires at
least one hit in each BBC arm whereas for $Au+Au$ at least two hits in
each BBC arm and one detected neutron in ZDC is needed. In order to
benefit from the high luminosity in \pp and \dau collisions and to
efficiently detect electrons, a special online EMCal RICH trigger
(ERT) is used. It requires an event to have at least one track with an
energy above a certain threshold in the EMCal and a geometrically
correlated hit in the RICH. The results presented here correspond to
the data for $p+p$ (2005), $d+Au$ (2003) and $Au+Au$ (2004) taken at
\sqn = 200 GeV \sqn = 62.4 GeV. 

\vspace{-0.3cm}
\section{Analysis Details}
\begin{figure}[!ht]
    \begin{center}
    \subfigure[\phk in Au+Au $@$ 200GeV]{
       \label{fig:fig_inv_spec_all-a} 
       \includegraphics[width=3in, height=2.35in]{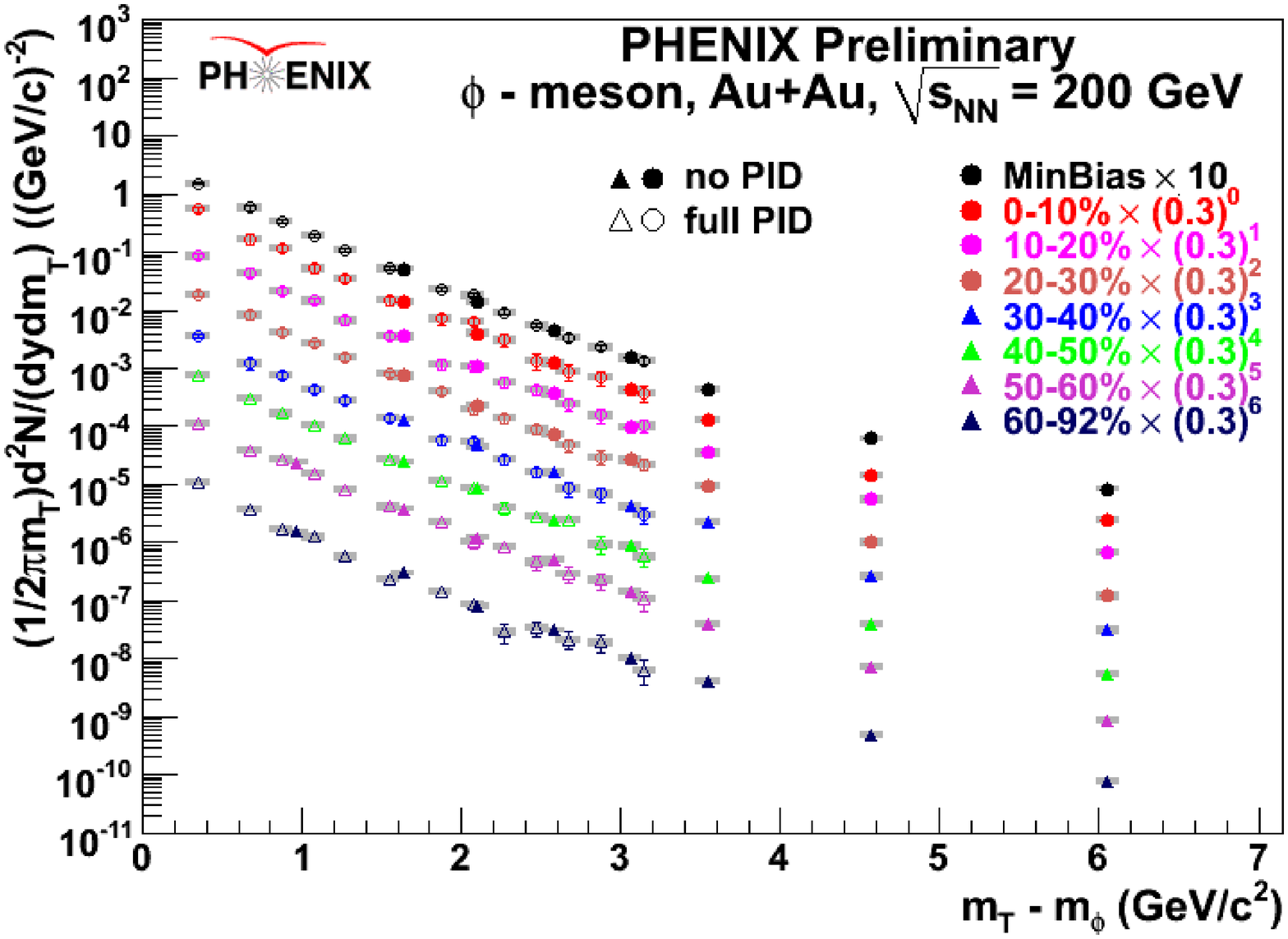}
     }
     \subfigure[\phe in Au+Au $@$ 200GeV]{
       \label{fig:fig_inv_spec_all-b}
       \includegraphics[width=2.9in]{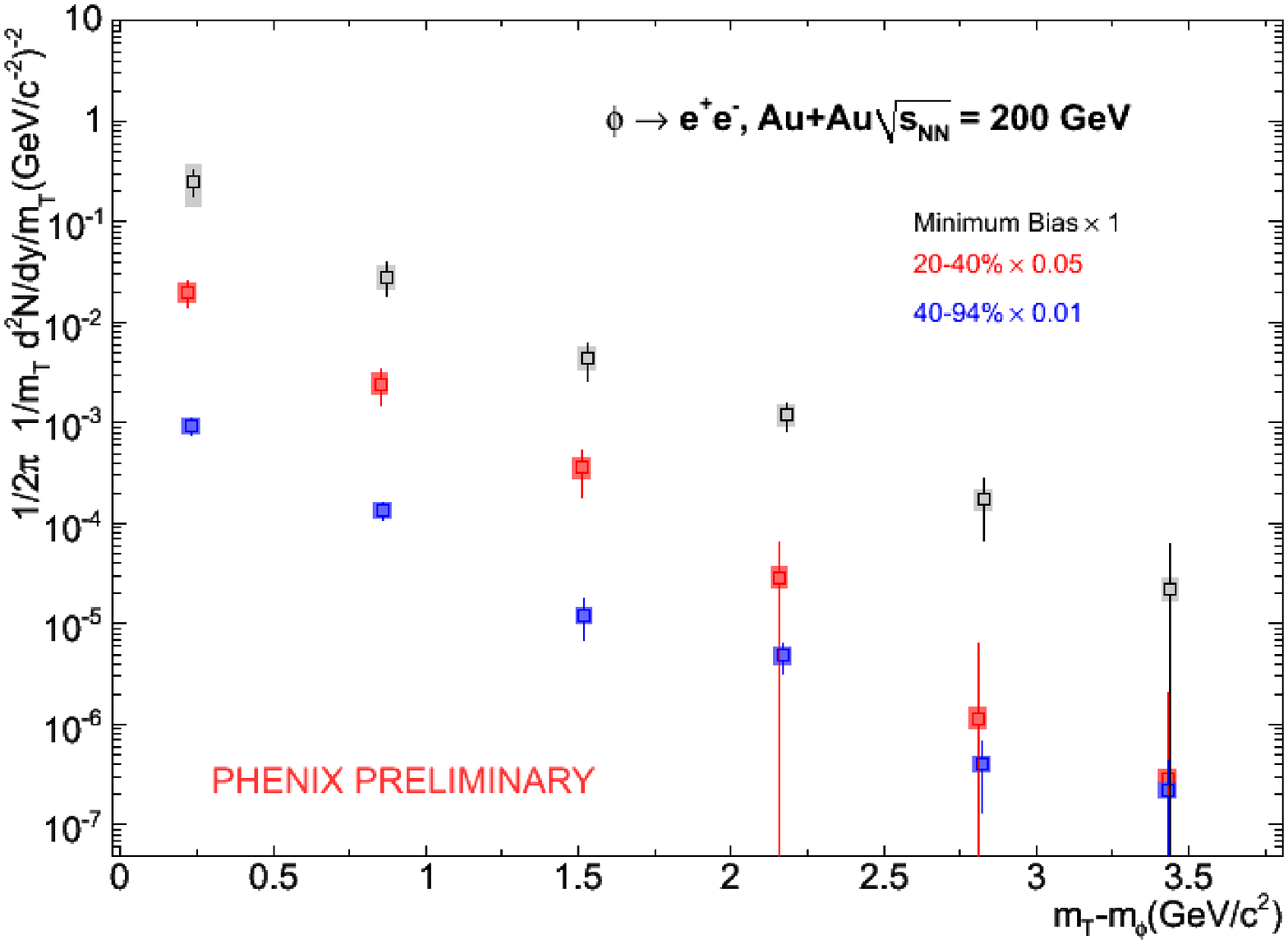}
     }
     \subfigure[$\phi \rightarrow K^{+}K^{-}$, $e^{+}e^{-}$, in p+p and
     d+Au $@$ 200GeV]{
       \label{fig:fig_inv_spec_all-c}
     \includegraphics[width=3.1in, height=2.35in]{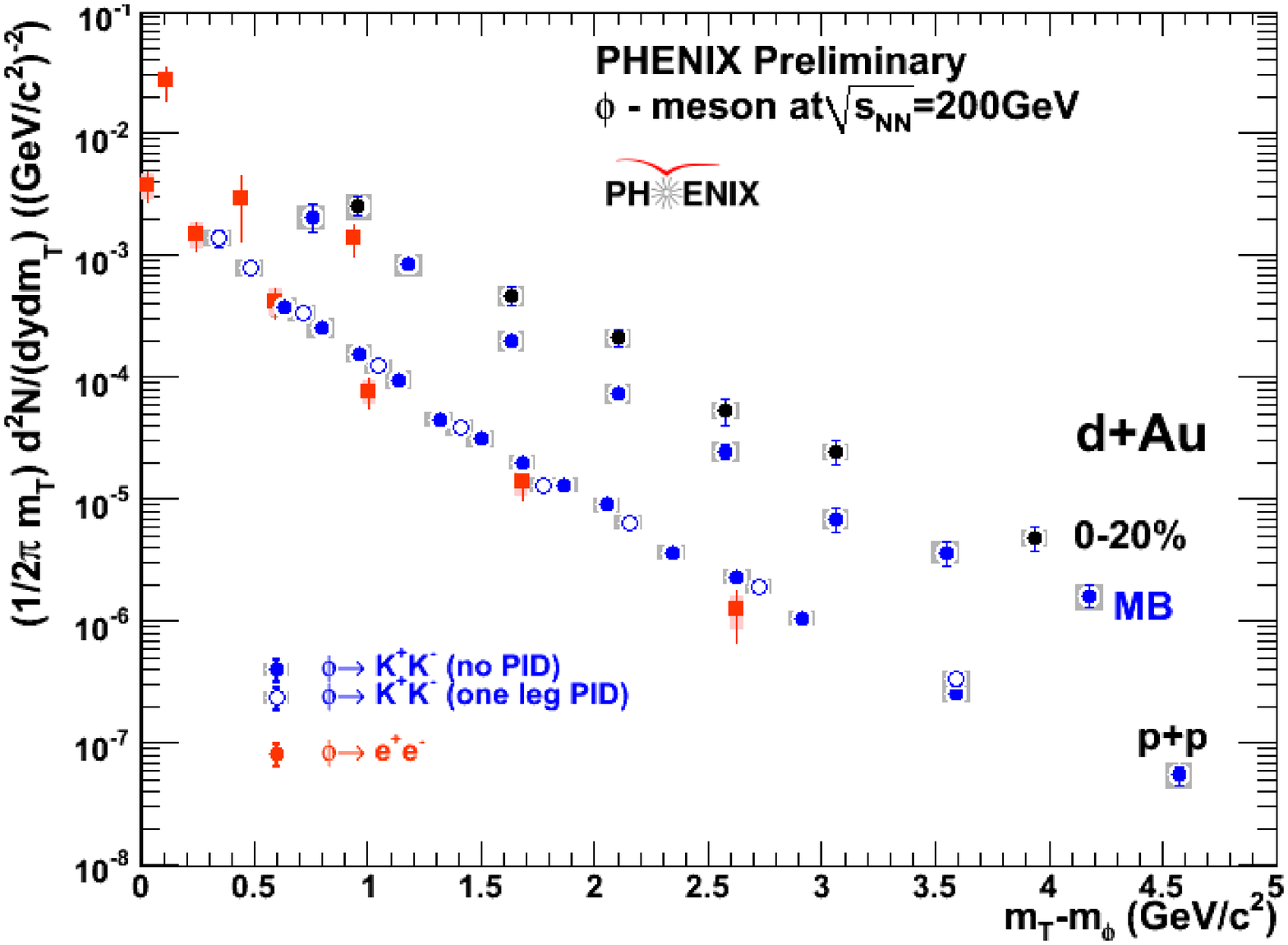}
   }
   \subfigure[\phk in Au+Au $@$ 62.4 GeV]{
     \label{fig:fig_inv_spec_all-d}
     \includegraphics[width=2.8in]{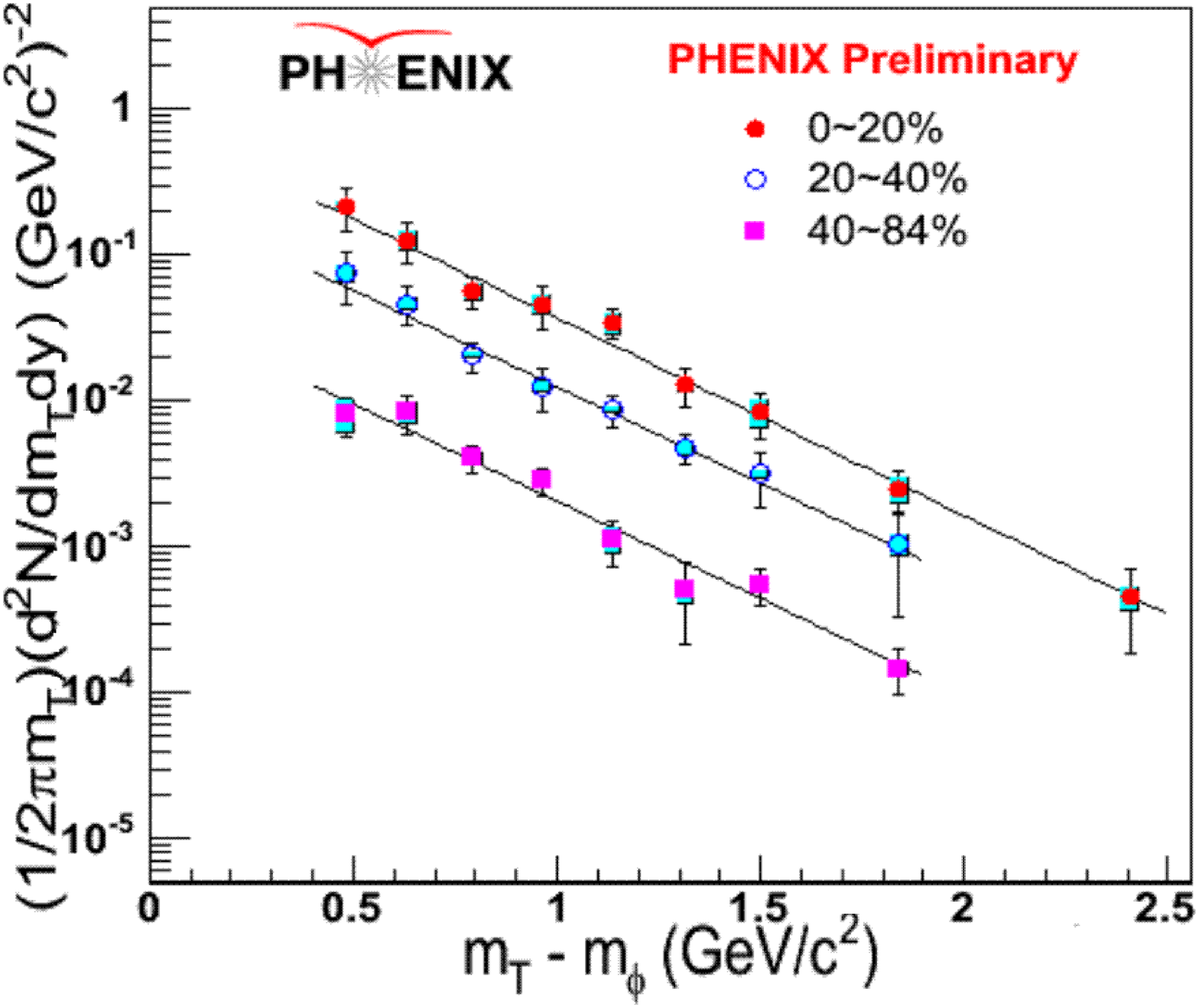}
   }
   \end{center}
    \vspace{-0.3cm}
    \caption{Invariant spectra of $\phi$-meson measured in different
      colliding systems.}
   \label{fig:fig_inv_spec_all}
 \end{figure}

For the \phk mass reconstruction, the charged tracks are combined to
form pairs using three different techniques. The first one does not
require any Kaon identification and assigns the kaon mass to all tracks
(``No Kaon PID''). The second method requires one track to be identified as
Kaon in the TOF subsystem (One Kaon``PID''), whereas the third method
requires both the tracks to be identified as Kaons either in TOF or
EMCal subsystems (``Two Kaon PID''). The \npid ~method allows us to go
to high $p_{T}$ as compared to the other two, is a relatively
simpler analysis, but with comparatively large B/S ratio. The \tpid
~method allows us to go lower in $p_{T}$ and has small B/S ratio. The
\opid ~method has the advantage of less background and so works with
better accuracy for the low $p_{T}$ ($<2$ GeV/\textit{c}) region where
\npid ~method has a large background. The \pp data was analyzed using
\npid, and \opid, \dau using \npid ~and \tpid~ and \au using \opid and
\tpid. The 62.4 GeV \au data was analyzed using \tpid ~method only. The
different analysis methods have very different sources of systematic
uncertainities and provide a valuable consistency check. In
Fig.~\ref{fig:fig_inv_spec_all-a}, good agreement between the various
methods can be seen. The combined \pp result using \opid ~and \npid
~analyses constitutes a new \pp reference for $\phi$-meson, surpassing
the previous one\cite{oldpp}, in $p_{T}$ and with smaller errors. 

For \phe, electrons identified using RICH and EMCal are combined in
pairs to generate like- and unlike-sign mass spectra. However, due to
the limited azimuthal angular acceptance and the strong magnetic field
beginning at R=0, the identification and rejection of $e^{+}e^{-}$
pairs from Dalitz decays and photon conversions is very
difficult\cite{kozlov2}. This results in a huge combinatorial
background in \au, making this measurement difficult.

Raw yields, for both $K^{+}K^{-}$ and $e^{+}e^{-}$ are then extracted
either by simultaneously fitting the signal and background, or by
integrating the spectra in the vicinity of the peak after subtraction
of the combinatorial background, estimated using an event-mixing
technique\cite{ryabov, kozlov}. The correction function takes into
account the limited detector acceptance and resolution and
reconstruction efficiency and was derived by a full single-particle
Monte Carlo simulation, propagated through an emulator of the PHENIX
detector and the full analysis chain. Corrections to incorporate the
trigger efficiency and multiplicity effects are applied too. 
Fig.~\ref{fig:fig_inv_spec_all} summarizes the $\phi$ measurements in
$p+p$, \dau and \au collisons via di-kaon and di-electron decay modes. 
\vspace{-1cm}
\section{Results}

\subsection{dN/dy and T}
\begin{figure}[!ht]
  \begin{center}
    \subfigure[``T'' as a function of system size and energy]{
      \label{fig:fig_dndy_t-a} 
      \includegraphics[width=2.9in]{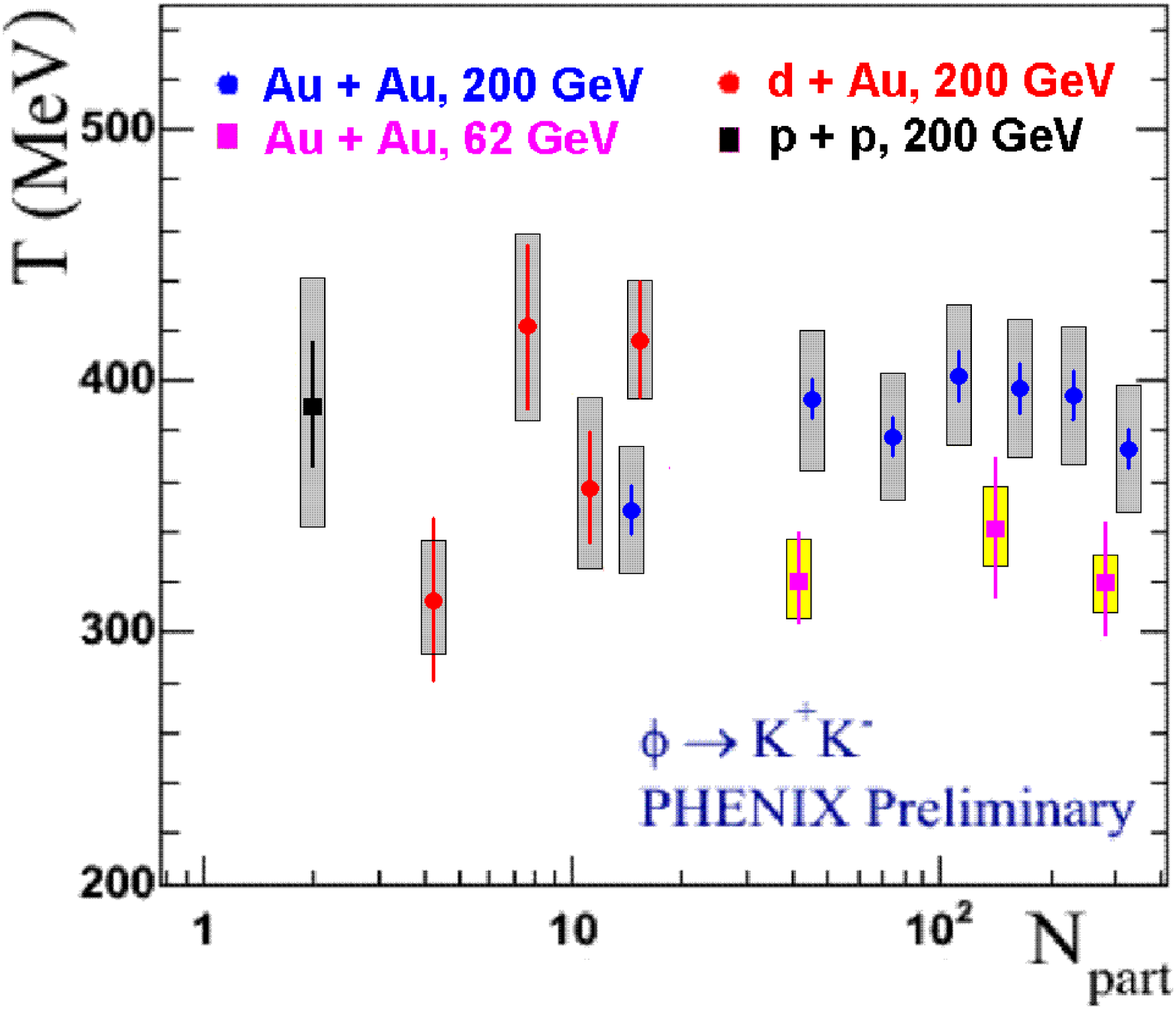}
    }
    \subfigure[Normalized yields as a fn. of $N_{part}$]{
      \label{fig:fig_dndy_t-b} 
      \includegraphics[height=2.5in,width=3in]{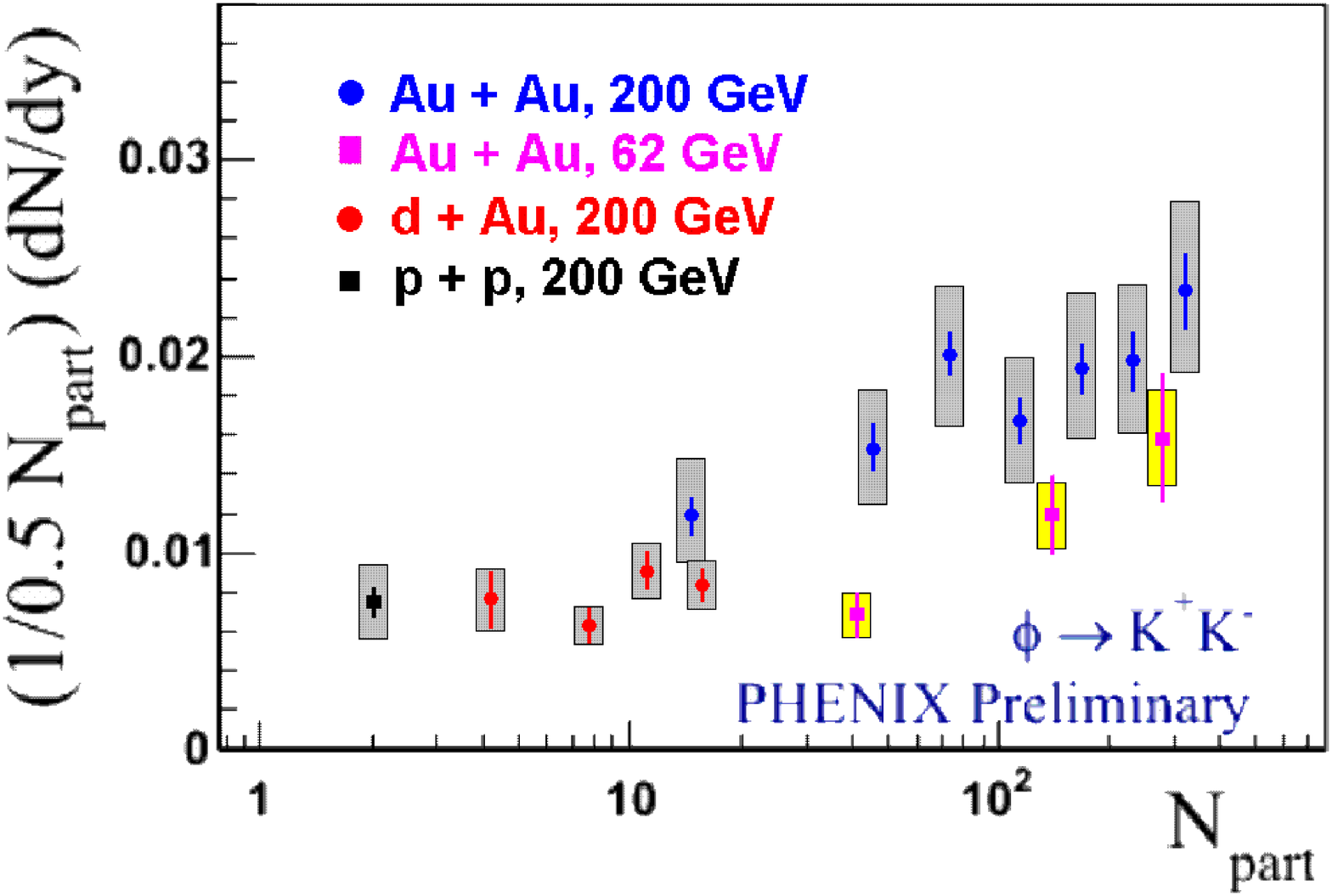}
    }
    \subfigure[``T'' comparison between $e^{+}e^{-}$ and $K^{+}K^{-}$]{
      \label{fig:fig_dndy_t-c} 
      \includegraphics[width=2.9in]{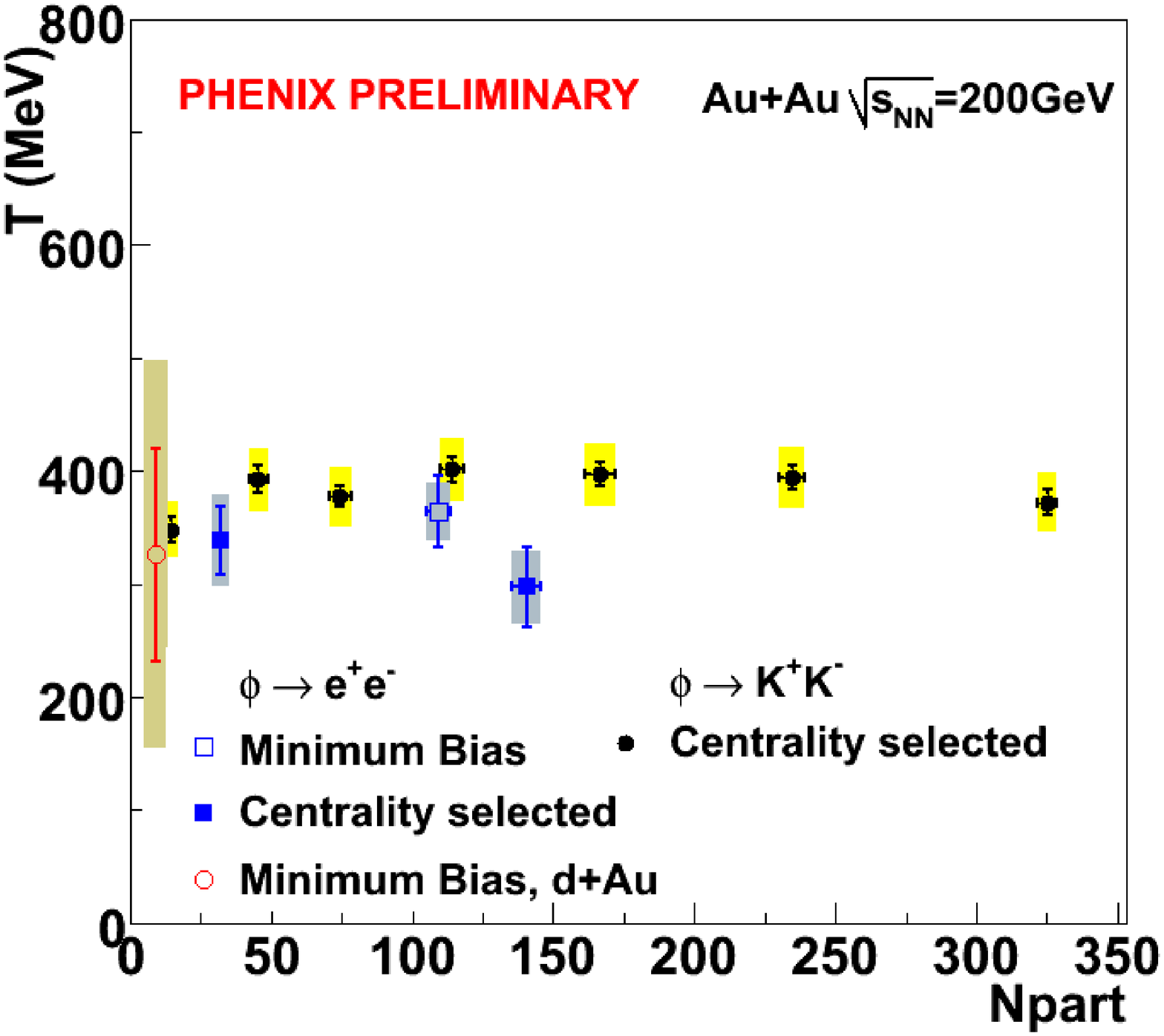}
      }
      \subfigure[dN/dy comparison between $e^{+}e^{-}$ and $K^{+}K^{-}$]{
      \label{fig:fig_dndy_t-d} 
      \includegraphics[width=3in]{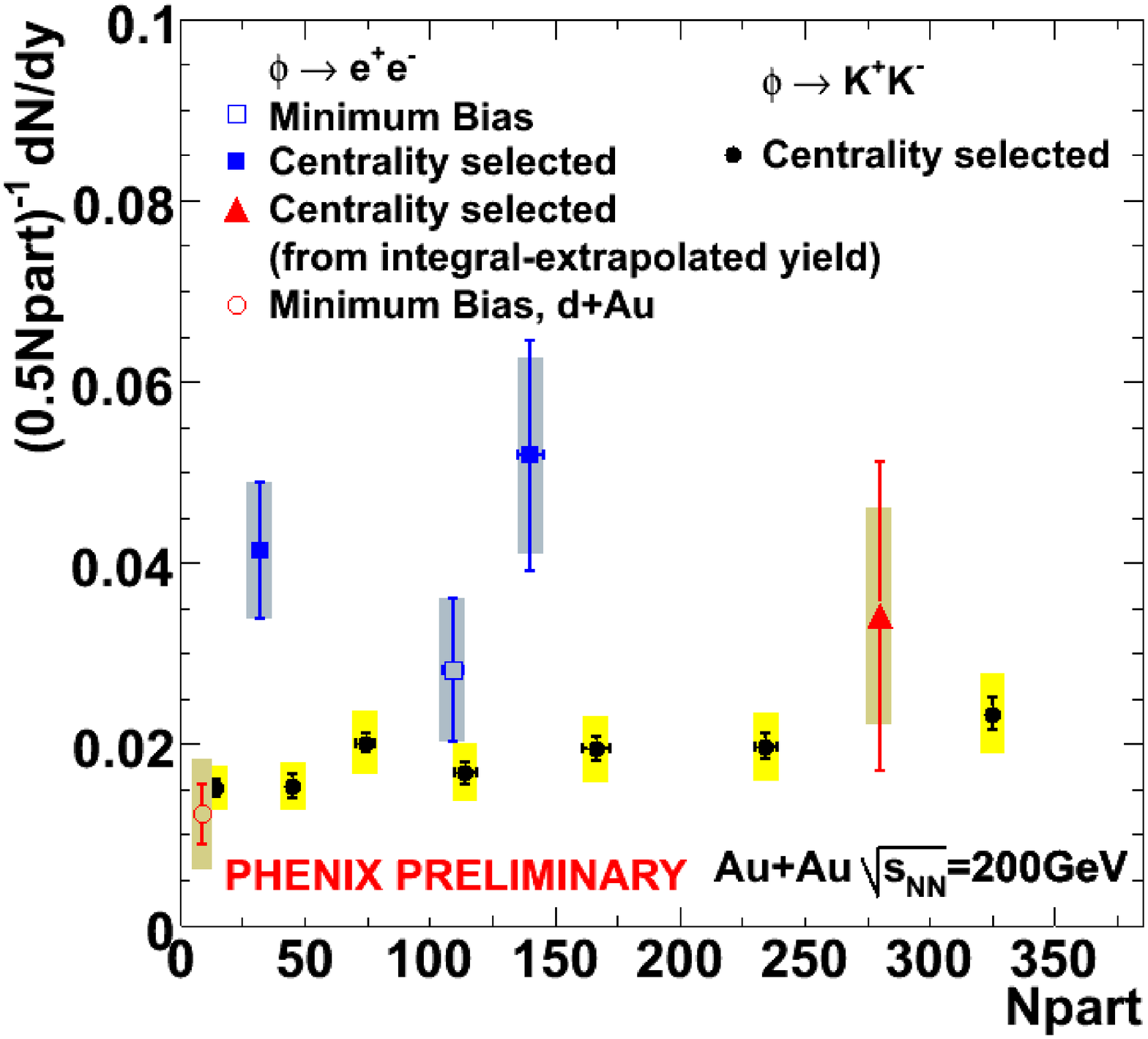}
    }
  \end{center}
  \vspace{-3mm}
  \caption{Integrated yields and Temperature}
  \label{fig:fig_dndy_t}
\end{figure}

The integrated yields and temperature presented in
Fig.~\ref{fig:fig_dndy_t} were extracted assuming $m_{T}$ dependent
yields to be exponential and so fitting the $m_{T}$ spectra to
the following exponential function with dN/dy and T as the
parameters.

\begin{equation}
  \frac{1}{2\pi m_T} \cdot \frac{d^{2}N} {dm_{T}dy}  = \frac {dN/dy}
  {2\pi T(T + M_{\phi})}  \cdot exp(-(m_{T} - M_{\phi})/T) 
  \nonumber
\end{equation}

The temperatures extracted in the two decay channels are in good
agreement with each other, do not change between different collision
systems but grow only slighlty from 62.4 GeV to 200 GeV as can be seen
in Figs.~\ref{fig:fig_dndy_t-a} ~\& ~\ref{fig:fig_dndy_t-c}. The
integrated yields normalized to the number of participants increase by
almost a factor of two from peripheral to central \au collisions. 
Particle yields in the $e^{+}e^{-}$ channel seem to be higher than the
corresponding $K^{+}K^{-}$ decay channel, but the large systematic and
statistical uncertainties in di-electron measurement prevent us from
making any conclusive statement.  

\subsection{$v_{2}$ of $\phi$}
\begin{figure}[!ht]
  \begin{center}
    \includegraphics[height=3.0in,width=\textwidth]{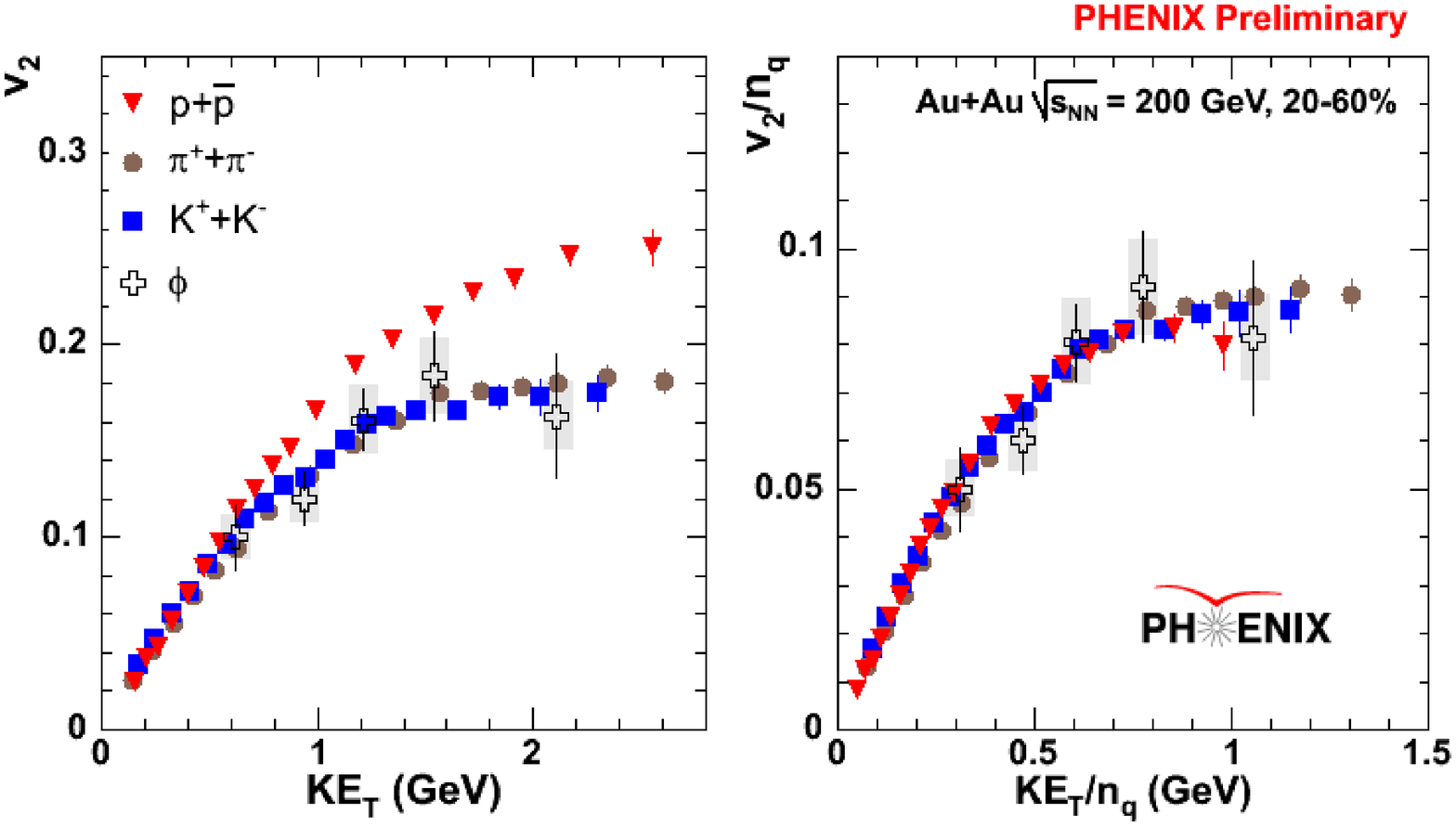} 
  \end{center}
  \caption{(a) $v_{2}$ vs $KE_{T}$ for identified particles\cite{chargev2} in
    mid-central (20-60\%) \au collisons. (b) $v_{2}/n_{q}$ vs $KE_{T}/n_{q}$
  for the same set of particles }
  \label{fig:fig_v2}
 \end{figure}

The $v_{2}$ of $\phi$ was extracted following the $m_{inv}$
method\cite{minv}, using the $K{+}K^{-}$ decay mode. A detailed
description of the analysis procedure can be found in \cite{phiv2}. 
Results for $v_{2}$ as a function of $KE_{T}$ for the $\phi$-meson 
together with $\pi^{\pm}, K^{\pm}~\&~(\overline{p})p$ are shown in
the left (right) panel of Fig.~\ref{fig:fig_v2} without (with) scaling
with the number of valence quarks. It is clear from the left panel
that despite its similar mass to the proton, $v_{2}(KE_{T})$ for the
$\phi$- meson follows the flow pattern of the other lighter
mesons($\pi$ and $K$). The right panel of the Fig.~\ref{fig:fig_v2}
shows universal scaling behavior of the elliptic flow of baryons and
mesons when both $v_{2}$ and $KE_{T}$ are scaled with the number of
valence quarks. This indicates that the elliptic flow develops
during the early stage when the constituents of the flowing medium are
partons and not ordinary hadrons interacting with their standard
hadronic cross-sections.

\subsection{Nuclear Modification Factor}
The nucleus-nucleus collisions at RHIC have revealed a decrease in
hadron production at high $p_{T}$ relative to scaled $p+p$
collisions. The nuclear medium effects on hadron production are
quantified by the nuclear modification factor $R_{AA}$, defined
as the ratio of the yield obtained from nucleus-nucleus collisions
scaled down with the number of binary collisions, to the yield from
elementary nucleon-nucleon collisions.

\begin{equation}
  R_{AA}(p_{T}) = \frac{d^{2}N^{AA}/dp_{T}dy} {\langle
    n_{coll} \rangle \cdot d^{2}N^{pp}/dp_{T}dy} \nonumber
\end{equation}

\begin{figure}[!ht]
  \begin{center}
    \includegraphics[width=0.48\textheight]{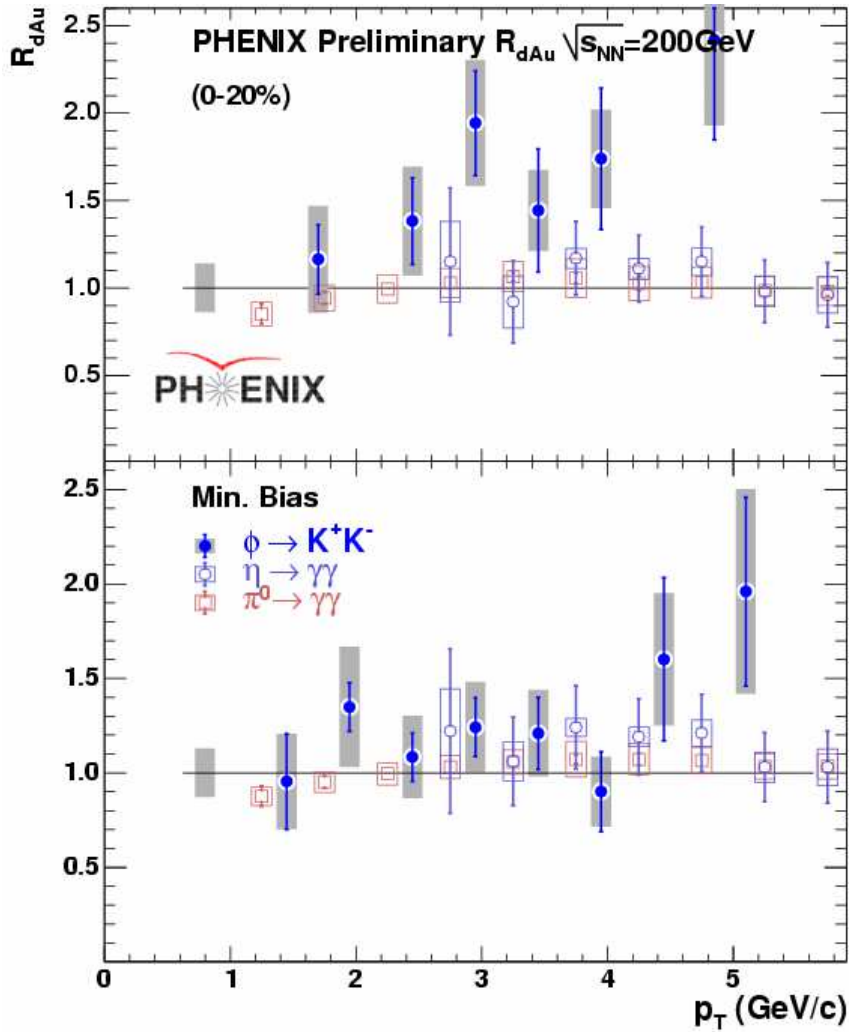}
  \end{center}
  \caption{$R_{dAu}$ of $\phi, \eta ~\& ~\pi^{0}$ mesons\cite{rda}.}
  \label{fig:fig_Rda}
\end{figure}

In the absence of any nuclear effects, the ratio should saturate at
unity for high $p_{T}$, where the production is dominated by hard
scattering and is proportional to the number of binary collisions
($N_{coll}$), in contrast to low $p_{T}$ region where soft processes
dominate. $R_{dAu}$, which is the nuclear modification factor for
$d+Au$ collisions helps to differentiate initial state effects from
final state effects and also to study other cold nuclear matter
effects. Fig.\ref{fig:fig_Rda} shows $R_{dAu}$ measurement of the $\phi$ for
0-20\% and minimum bias collisions plotted together with $\eta
\rightarrow \gamma\gamma$ and $\pi^{0} \rightarrow \gamma\gamma$
mesons for comparison. Whereas the $R_{dAu}$ of $\pi$ and $\eta$ are
consistent with unity, the $R_{dAu}$ of $\phi$ for 0-20\% shows some
enhancement. but with very large error bars.


\begin{figure}[!ht]
  \begin{center}
    \includegraphics[width=4in, angle=90]{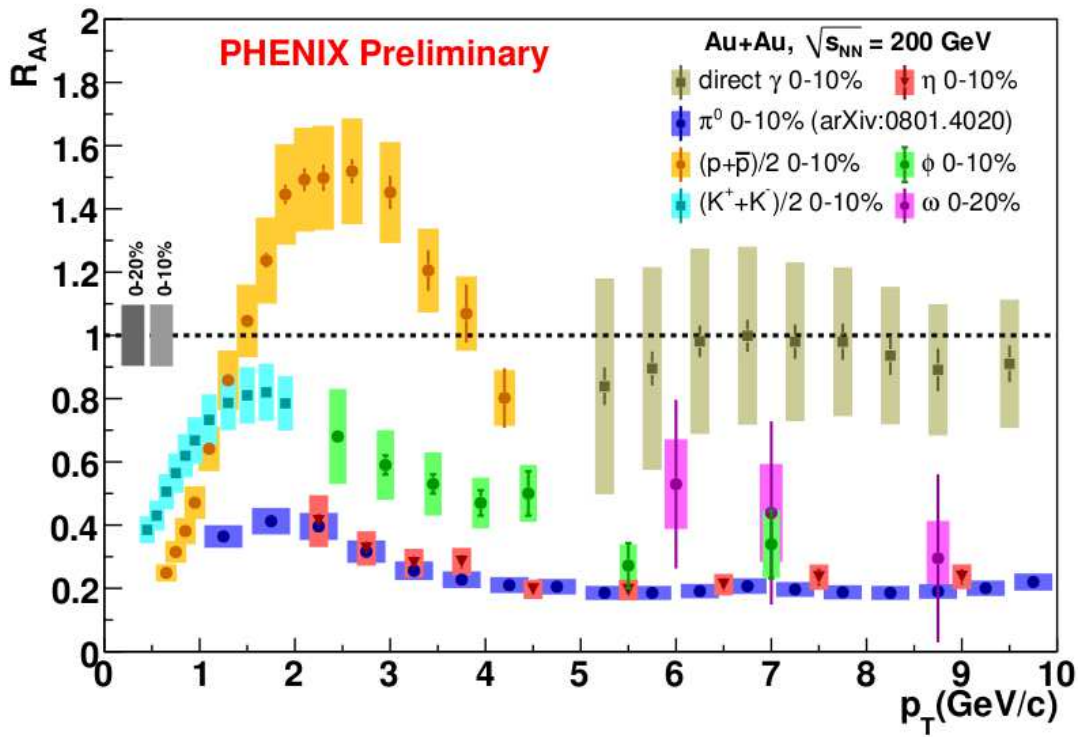}
  \end{center}
  \caption{Nuclear modification factor $R_{AA}$ in central \au
    collisions as a function of $p_{T}$ for $\pi^{0}, (K^{+} +
    K^{-})/2, \eta, \omega, (p + \overline{p})/2, \phi$ and direct
    $\gamma$.} 
  \label{fig:fig_RAA}
\end{figure}

Fig.\ref{fig:fig_RAA}\cite{maxim} shows the nuclear modification
factor for the $\phi$-meson in central \au collisions using the new
$p+p$ reference from 2005 run results, over a $p_{T}$ range of 2.45 - 7
GeV/\textit{c}. For comparison, the $R_{AA}$ of direct $\gamma, \pi^{0}, \eta, \omega,
(K^{+} + K^{-})/2 ~\& ~(p + \overline{p})/2$ are also plotted. $\pi^{0}$
and $\eta$ follow the same suppression pattern over the entire $p_{T}$
range. The $\phi$-meson at intermediate $p_{T}$($2.45 < p_{T} < 4.5 $ GeV/\textit{c})
exhibits more suppression than the protons but less suppression than
$\eta$ and $\pi^{0}$ mesons, whereas at higher $p_{T}$ ($> 5$
GeV/\textit{c})
the amount of suppression of $\phi$ and $\omega$ could be similar to
that of $\pi^{0}$ and $\eta$. The similar suppression patterns of
various mesons at high $p_{T}$ support the concept of meson production
via jet-fragmentation at high $p_{T}$ outside the hot and dense medium
created in the collisions. In the future, the comparison of extended measurements of
charged Kaons with that of $\phi$ will give an insight into, whether
or not quark flavor composition plays a role on the suppression pattern.

\section{Summary}
The PHENIX experiment at RHIC has measured $\phi$-meson production
using both $e^{+}e^{-}$ and $K^{+}K^{-}$ in \pp, \dau and \au
collisions at \sqn = 200 GeV. The preliminary results in \pp via both
channels agree to each other. The measurements in \dau suffer from low
statistics in both the channels. With the high statistics \dau data
collected in year 2008, results with better precision and wide $p_{T}$
coverage are expected. In \au, the $\phi$-meson has been measured via $K^{+}K^{-}$ with
good accuracy while $e^{+}e^{-}$ measurements are expected to improve
considerably with the newly installed Hadron Blind Detector\cite{hbd,
nim1, nim2} in PHENIX. The $v_{2}$ of $\phi$ follows the flow
pattern of other 
mesons as well as scales with $n_{q}$ =2, indicating partonic
collectivity. $R_{dAu}$ of $\phi$ does not show any suppression, but
large error bars leave room for Cronin enhancement. The $R_{AA}$ of
$\phi$ shows a similar suppression pattern to that of $\pi^{0}$ and
$\eta$ at high $p_{T}$, but at intermediate $p_{T}$, the suppression
level is different.

\section{Acknowledgements }
The author acknowledges support by the Israel Science Foundation, the
MINERVA Foundation and the Nella and Leon Benoziyo Center of High
Energy Physics Research.

\end{document}